\documentstyle[12pt]{article}
\newcommand{\nc}{\newcommand}

\nc{\dbar}{\bar{\partial}}
\nc{\be}{\begin{equation}}
\nc{\ee}{\end{equation}}

\catcode`\@=11

\@addtoreset{equation}{section}
\def\theequation{\thesection\arabic{equation}}

\def\@normalsize{\@setsize\normalsize{15pt}\xiipt\@xiipt
\abovedisplayskip 14pt plus3pt minus3pt%
\belowdisplayskip \abovedisplayskip
\abovedisplayshortskip  \z@ plus3pt%
\belowdisplayshortskip  7pt plus3.5pt minus0pt}
\def\small{\@setsize\small{13.6pt}\xipt\@xipt
\abovedisplayskip 13pt plus3pt minus3pt%
\belowdisplayskip \abovedisplayskip
\abovedisplayshortskip  \z@ plus3pt%
\belowdisplayshortskip  7pt plus3.5pt minus0pt
\def\@listi{\parsep 4.5pt plus 2pt minus 1pt
            \itemsep \parsep
            \topsep 9pt plus 3pt minus 3pt}}

\def\underline#1{\relax\ifmmode\@@underline#1\else
        $\@@underline{\hbox{#1}}$\relax\fi}
\@twosidetrue
\relax

\catcode`@=12

\catcode`\@=11

\def\section{\@startsection{section}{1}{\z@}{3.5ex plus 1ex minus
   .2ex}{2.3ex plus .2ex}{\large\bf}}
\def\thesection{\arabic{section}.}

\def\ps@headings{\def\@oddfoot{}\def\@evenfoot{}
\def\@oddhead{\hbox{}\hfill
        \makebox[.5\textwidth]{\raggedright\ignorespaces --\thepage{}--
        \hfill }}
\def\@evenhead{\@oddhead}
\def\subsectionmark##1{\markboth{##1}{}}
}

\ps@headings

\catcode`\@=12

\relax

\def\figcap{\section*{Figure Captions\markboth
        {FIGURECAPTIONS}{FIGURECAPTIONS}}\list
        {Fig. \arabic{enumi}:\hfill}{\settowidth\labelwidth{Fig. 999:}
        \leftmargin\labelwidth
        \advance\leftmargin\labelsep\usecounter{enumi}}}
 \relax
\def\tablecap{\section*{Table Captions\markboth
        {TABLECAPTIONS}{TABLECAPTIONS}}\list
        {Table \arabic{enumi}:\hfill}{\settowidth\labelwidth{Table 999:}
        \leftmargin\labelwidth
        \advance\leftmargin\labelsep\usecounter{enumi}}}
 \relax
\def\reflist{\section*{References\markboth
        {REFLIST}{REFLIST}}\list
        {[\arabic{enumi}]\hfill}{\settowidth\labelwidth{[999]}
        \leftmargin\labelwidth
        \advance\leftmargin\labelsep\usecounter{enumi}}}
 \relax

\catcode`\@=11

\def\ps@headings{\def\@oddfoot{}\def\@evenfoot{}
\def\@oddhead{\hbox{}\hfill
        \makebox[.5\textwidth]{\raggedright\ignorespaces --\thepage{}--
        \hfill }}
\def\@evenhead{\@oddhead}
\def\subsectionmark##1{\markboth{##1}{}}
}

\ps@headings

\relax

\def\firstpage#1#2#3#4#5#6{
\begin{document}

\begin{titlepage}
\nopagebreak
\title{\begin{flushright}
        \vspace*{-1.8in}
        {\normalsize IC/97/24}\\[-9mm]
        {\normalsize hep-th/9704006}\\[4mm]
\end{flushright}
\vfill
{\large \bf #3}}
\author{\large #4 \\ #5}
\maketitle
\vskip -7mm
\nopagebreak
\begin{abstract}
{\noindent #6} 
\end{abstract}
\vfill
\begin{flushleft}
\rule{16.1cm}{0.2mm}\\[-3mm]
$^{\star}${\small Research supported in part by the EEC contract \vspace{-4mm}
SCI-CT920792.}\\
March 1997
\end{flushleft}
\thispagestyle{empty}
\end{titlepage}}
\newcommand{\dal}{\raisebox{0.085cm}
{\fbox{\rule{0cm}{0.07cm}\,}}}
\newcommand{\dt}{\partial_{\langle T\rangle}}
\newcommand{\dtbar}{\partial_{\langle\bar{T}\rangle}}
\newcommand{\al}{\alpha^{\prime}}
\newcommand{\mst}{M_{\scriptscriptstyle \!S}}
\newcommand{\mpl}{M_{\scriptscriptstyle \!P}}
\newcommand{\dv}{\int{\rm d}^4x\sqrt{g}}
\newcommand{\lv}{\left\langle}
\newcommand{\rv}{\right\rangle}
\newcommand{\ph}{\varphi}
\newcommand{\sbar}{\,\bar{\! S}}
\newcommand{\xbar}{\,\bar{\! X}}
\newcommand{\fbar}{\,\bar{\! F}}
\newcommand{\zbar}{\,\bar{\! Z}}
\newcommand{\tbar}{\bar{T}}
\newcommand{\ubar}{\bar{U}}
\newcommand{\ybar}{\bar{Y}}
\newcommand{\phb}{\bar{\varphi}}
\newcommand{\cm}{Commun.\ Math.\ Phys.~}
\newcommand{\pr}{Phys.\ Rev.\ D~}
\newcommand{\pl}{Phys.\ Lett.\ B~}
\newcommand{\prl}{Phys.\ Rev.\ Lett.~ }
\newcommand{\ibar}{\bar{\imath}}
\newcommand{\jbar}{\bar{\jmath}}
\newcommand{\np}{Nucl.\ Phys.\ B~}
\newcommand{\e}{{\rm e}}
\newcommand{\gsi}{\,\raisebox{-0.13cm}{$\stackrel{\textstyle
>}{\textstyle\sim}$}\,}
\newcommand{\lsi}{\,\raisebox{-0.13cm}{$\stackrel{\textstyle
<}{\textstyle\sim}$}\,}
\date{}
\firstpage{95/XX}{3122}
{\large\sc  On the Bound States of $p$- and $(p+2)$-Branes} 
{E. Gava$^{a,b}$, 
K.S. Narain$^{ b}$ $\,$and$\,$
M.H. Sarmadi$^{\,b}$}
{\normalsize\sl
$^a$Instituto Nazionale di Fisica Nucleare, sez.\ di Trieste,
Italy\\[-3mm]
\normalsize\sl $^b$International Centre for Theoretical Physics,
I-34100 Trieste, Italy\\[-3mm]}
{We study bound states of D-$p$-branes
and D-$(p+2)$-branes. By switching on a large magnetic field $F$
on the $(p+2)$ brane, the problem is shown to admit a perturbative analysis
in an expansion in inverse powers of $F$. It is found that, to 
the leading order in $1/F$, the quartic potential of the 
tachyonic state from the open string stretched between the $p$- and
$(p+2)$-brane gives a vacuum energy which agrees 
with the prediction of the BPS mass formula
for the bound state. We generalize the discussion to the case of
$m$ $p$-branes plus 1 $(p+2)$-brane with magnetic field. The $T$ dual
picture of this, namely several $(p+2)$-branes carrying some $p$-brane
charges through magnetic flux is also discussed, where the perturbative
treatment is available in the small $F$ limit. We show that once again,
in the same approximation, the
tachyon condensates give rise to the correct BPS mass formula.
The role of 't Hooft's toron configurations in the
extension of the above results beyond the quartic approximation as
well as the issue of the unbroken gauge symmetries are discussed. 
We comment on the
connection between the present bound state problem and Kondo-like problems
in the context of relevant boundary perturbations of boundary conformal
field theories.}

\section{Introduction}

The question of bound states of different p-branes have played an important
role in the understanding of various dualities. For example, the 
$SL(2,Z)$ duality symmetry of type IIB, predicts bound states of 
$n$ number of 1-branes carrying R-R and $m$ number of the ones carrying
NS-NS charge with $n$ and $m$ being relatively prime \cite{sc,wit}. 
Indeed a 1-brane 
carrying the above charges is expected to have a tension given by the
BPS expression:
\begin{equation}
{\rm T}(n,m)=T \sqrt{\frac{n^2}{\lambda^2}+m^2}
\label{tension}
\end{equation}
where $\lambda$ is the string coupling constant and $T=\frac{1}{2\pi 
\alpha'}$. Throughout we will set $\alpha'=1/2$.
Since ${\rm T}(n,m)$ is less than the sum of 
the tensions of individual 1-branes,
ie. $T(n/\lambda+m)$, this system is expected to form a bound state.
This was shown by Witten \cite{wit} by 
considering the world-sheet theory of $n$ 
D-branes \cite{polD} which is a supersymmetric $N=8$, $U(n)$ gauge theory.
 The NS-NS 
charge is introduced by giving vacuum expectation value to field strengths
that correspond to a source transforming as $m$-th rank anti-symmetric 
tensor representation of $U(n)$. As a result, the $SU(n)$ part of $U(n)$
develops a mass gap showing the formation of a bound state, while the 
$U(1)$ part which corresponds to the center of mass motion gives rise to
the same degrees of freedom as that of the fundamental string. The above 
expression for the tension follows from the Born-Infeld action for the 
center of mass $U(1)$ taking into account the above expectation value of
its field strength. This is in agreement with the fact that the fundamental 
string is in the same $SL(2,Z)$ orbit as $(n,m)$ string.   

Starting from this result and applying various $T$ and $S$-dualities one
can arrive at statements about the existence of bound states of other
p-brane systems. For example let us compactify two of the transverse 
directions of the $(n,m)$- string on a torus. Performing two $T$ duality
transformations one gets a system of $n$ D-3-branes carrying $m$ units of
NS-NS 1-brane charge through the non-zero expectation value of the 
electric field. Further application of $S$-duality 
turns this system to that of $n$ D-3-branes with $m$ units of D-1-brane 
charge. As a result of $S$-duality transformation the original non-zero 
expectation value of electric field is transformed to a non-zero magnetic
field on the torus. This is in agreement with the existence of the 
Chern-Simons term $\int F\wedge A_2$ in the 3-brane world volume action
where $A_2$ is the R-R 2-form. As noted in \cite{D}, the presence of this term
implies that, when $F$ has non trivial expectation value, the 3-brane carries
D-1-charge.
In this picture the quantization of 
D-1-charge is just a consequence of the quantization of the magnetic 
flux on the torus. One can easily see that under these duality 
transformations the tension given in eq. (\ref{tension}) is transformed to the
following expression:
\begin{equation}
{\rm T}(n,m)= \frac{TV}{2\pi^2\lambda}\sqrt{n^2 +\frac{4\pi^4m^2}{V^2}}
\label{tension'}
\end{equation}
where $V$ is volume of the torus. Note that although in the above discussion 
we restricted ourselves to 3- and 1-brane system, by $T$-duality it also 
applies to a system of $(p+2)$- and $p$-branes.
In the above approach the $p$-brane charge appeared due to the expectation 
value of the magnetic field on the $(p+2)$-brane. However one can ask the 
question what happens when there are separate $(p+2)$- and $p$-branes. It is
known that this system is not supersymmetric due to the number of mixed
Neumann-Dirichlet
(ND)-directions being 2 \cite{pol}. Moreover the lowest mode of the string 
stretched between the $p$- and $(p+2)$- branes has the mass given by:
\begin{equation}
m^2 = -\frac{1}{2} +\frac{b^2}{\pi^2}
\label{mass}
\end{equation}
where $b$ is the distance between the two branes. As a result when 
$b^2 \le \frac{\pi^2}{2}$ the above mode becomes tachyonic. One expects
\cite{pol} that the dynamics of this tachyon will give a ground state
in accordance with what we saw above from the application of dualities
on $(m,n)$-string \cite{PS}. In particular, since the ground state 
energy of a system of separated $p$- and $(p+2)$- branes
is strictly greater than the BPS tension for the bound state, 
the tachyon should provide a negative
contribution to the vacuum energy of the  $p-(p+2)$ system
by the amount given  by the difference between the bound state
tension and the sum of the individual tensions.

In this paper we will study various configurations of $p$ and
$(p+2)$-branes, where, considering the tachyon potential to the quartic order,
we show that to the
leading approximation in small (or large) volume limit 
the above expectation turns out to be correct. In
section 2, we study the tachyon potential when a $(1,N)$ system (ie. a
single $(p+2)$-brane carrying $N$ units of $p$-brane charge) is brought near
a single $p$-brane (ie. a $(0,1)$ system).
In Appendix we give the string derivation of the quartic potential
of the tachyon. 
We also generalize this result when
$(1,N)$ system is brought near $(0,m)$ system consisting of m $p$-branes. In
Section 3, we discuss the $T$-dual situation of the previous section. Thus
we consider $N$ $(p+2)$-branes where one of the branes carries a $p$-brane
charge via a unit magnetic flux and arrive at the bound state formation by
minimzing the tachyon potential. 
In that section we
also discuss what happens when an $(N,1)$ system is brought near a (1,0)
system, which is the exact $T$-dual of the situation considered in section
2. In section 4, we discuss $(N,m)$ bound states obtained by taking $N$ 
$(p+2)$-branes with $m$ units of magnetic fluxes. In this case again we find
bound states. In section 5, we show that the above results can be extended
beyond the quartic order and the resulting solution
turns out to be related to 't Hooft's torons \cite{thooft}.
We also discuss the issue of the unbroken gauge
symmetries in the presence of these torons and verify that
the massive Kaluza-Klein spectrum agrees with $T$-duality.
In Section 6, we make some
concluding remarks and in particular show that string duality implies the
existance of non-trivial fixed points in a class of boundary conformal field
theories that are perturbed by some relevant boundary operators. In fact the
cases considered in the present paper are just the situations when the
relevant operators are almost marginal. However the string duality predicts
that this phenomenon must continue to hold even when the relevant operators
are far from being marginal and there is a precise prediction on the value
of the $g$-function (ie. disc partition function) at the new fixed point.

\section{Bound State of $(1,N)$ and $(0,1)$ systems for large $N/V$}

Let us consider an $(1,N)$ system where $N$ units of $p$-brane charge 
has been introduced by turning on $N$ units of the
magnetic flux on the torus. Now imagine bringing a single $p$-brane close
to the $(1,N)$ system. The sum of the two tensions is 
\begin{equation}
{\rm T}(1,N)+{\rm T}(0,1)=
\frac{TV}{2\pi^2\lambda}(\sqrt{1+\frac{4\pi^4 N^2}{V^2}} + 
\frac{2\pi^2}{V}). 
\label{sum}
\end{equation}
On the
other hand the combined system which carries the charge $(N+1,1)$ should
have a tension 
\begin{equation}
{\rm T}(1,N+1)=\frac{TV}{2\pi^2\lambda}\sqrt{1+\frac{4\pi^4 
(N+1)^2}{V^2}}.
\label{bs}
\end{equation}
The first point to notice is the fact that the two tensions differ at order 
$1/\lambda$ and therefore one should be able to understand the
formation of the bound state by minimizing the tree level potential for 
the tachyons. The potential at the minimum should account for the
difference between the two tensions. 
For large $N/V$, the difference, $\delta {\rm T}={\rm T}(1,N+1)-{\rm T}(1,N)
-{\rm T}(0,1)$, between 
these two values of tension goes as:
\begin{equation}
\delta{\rm T}=\frac{-TV^2}{8\pi^4\lambda N(N+1)}+{\cal O}(\frac{V^3}{N^3}), 
\label{delta}
\end{equation}
and therefore one may hope to
find a solution in perturbation theory. This limit can be achieved either by
taking small volume limit for a fixed $N$, or by taking large $N$ limit for
a fixed volume. As we shall see below, the minimization of the tachyon 
potential to the quartic order reproduces the difference (\ref{delta}).

As mentioned above, the tachyon field is the lowest mode of the open 
string stretched between the $p$- and $(p+2)$- branes. On the $p$- brane 
side the string satisfies Dirichlet boundary condition, while on the $(p+2)$-
brane side Neumann condition.
In ref.\cite{call,bp,b}, the quantization of 
open strings in the presence of constant
magnetic field has been discussed. The effect of the magnetic field on 
the $(p+2)$-brane can 
be incorporated by modifying the boundary condition of the open string 
from Neumann to a mixed boundary condition given by
\begin{equation}
\partial_{\sigma} X =- i\pi F\partial_{\tau} X
\label{bcnone}
\end{equation}
where $X$ is the complex coordinate on the torus and $F=\frac{2\pi N}{V}$ is 
the quantized magnetic field.  
Taking into account the Dirichlet boundary condition on the $p$-brane 
side one has the following mode expansion for $X$:
\begin{eqnarray}
X&=& i\sum_{n=1}^{\infty}a_n \phi_n(\sigma,\tau)-i
\sum_{n=0}^{\infty}\tilde{a}^{\dagger}_n \phi_{-n}(\sigma,\tau) \nonumber \\
\phi_n(\sigma,\tau)&=&(n-\frac{1}{2}-\epsilon)^{-\frac{1}{2}}
e^{-i(n-\frac{1}{2}-\epsilon)\tau}cos[(n-\frac{1}{2}-\epsilon)\sigma
+\pi \epsilon)]
\label{x}
\end{eqnarray}
where
\begin{equation}
\epsilon = \frac{1}{\pi}arctan(\pi F)
\label{epsilon}
\end{equation}
There is also similar mode expansion for fermions. Note that in the above 
mode expansion we do not have a zero mode for $X$ due to the DN boundary 
condition. This is to be contrasted with the NN case studied in ref.
\cite{call,bp,b}
where $X$ has a zero mode $x$ which satisfy a modified commutation 
relation $[x,\bar{x}]= 1/F$ due to the presence of magnetic field. This 
resulted in a degeneracy of states $FV/2\pi=N$ due to the phase space factor.
In the present case since there is no zero mode of $X$, we do not have
this degeneracy. This can also be understood by starting from NN string
with magnetic field on both sides with fluxes being $M$ and $N$ units
respectively. The multiplicity of states in this case will be $M+N$. Now
taking $M\rightarrow \infty$ one arrives at the DN string that we are 
considering, but with a multiplicity factor which goes to infinity. This
multiplicity factor is just the choice of the position of the D-end of 
the string, which in our problem is fixed to be the position of the 
$p$-brane.  

Taking into account the zero point energy of the bosons and fermions,
one finds that the lowest modes of the states in the Neveu-Schwarz sector,
when the $p$-brane is inside the $(p+2)$-brane (ie. $b=0$) have masses:
\begin{equation}
m^2 = (2n-1)(\frac{1}{2}-\epsilon)
\label{mass'}
\end{equation}
Here $n$ labels the Landau levels obtained by applying 
$\tilde{a}^{\dagger}_0$. The tachyon of course appears only at the lowest 
Landau level ie. $n=0$. Now let us take large $N/V$ limit. From 
eq(\ref{epsilon}) it follows that $\epsilon \rightarrow 
\frac{1}{2}-\frac{V}{2\pi^3 N} + {\cal O}(\frac{V^2}{N^2})$. 
The mass square of 
the tachyon to the leading order is therefore $-V/2\pi^3 N$. 

Let us now discuss the quartic interaction term in the tachyon potential.
Since in the
large $N/V$ limit, the Neumann boundary condition on the $(p+2)$-brane end
becomes effectively Dirichlet, the string stretched between the two branes
becomes DD. This means that in this limit the tachyon vertex operator
which involves twist fields for bosons and fermions (twisting by 
$\frac{V}{2\pi^3N}$) goes over to that of the DD- scalar $\partial_{\sigma} X$.
We expect therefore that the quartic term in the
potential starts from order $(\frac{V}{N})^0$. If there is a non-zero
contribution to the quartic term at this leading order, then minimizing the
potential with respect to the tachyon field shows that its vacuum
expectation value (vev) is of order $\sqrt{\frac{V}{N}}$ and, correspondingly,
the vacuum energy is of order
$(\frac{V}{N})^2$: we will see that in fact it agrees with the BPS
prediction (\ref{delta}). 
The quartic term can be calculated
directly at the
string level by inserting four twist fields corresponding to the tachyon
fields on a disc and removing the reducible diagrams arising from the
exchange of the two $U(1)$ gauge fields. One can then take the limit $V/N
\rightarrow 0$ in which case the on-shell momenta of the tachyons satisfy
the massless condition and the potential can be calculated by taking the zero
momentum limit. We have given the details of this computation in the
Appendix.  Denoting the tachyon field by $\chi$, the resulting
potential upto quartic term (and upto the relevant order in $V/N$) is given
by: \begin{equation}
{\cal V} = -\frac{V}{2\pi^3N}\chi\bar{\chi} + 
\frac{1}{2\pi^2}(1+\frac{1}{N})(\chi\bar{\chi})^2 
\label{potential}
\end{equation}
which has a minimum at $\chi\bar{\chi} = \frac{V}{2\pi (N+1)}$. The resulting 
value
of the potential at this minimum is $-V^2/{8\pi^4 N(N+1)}$. This is 
exactly
the desired value to this order as we saw above when we considered the
difference $\delta{\rm T}$ between the 
tension of the bound state and the sum of the
individual ones given in eq.(\ref{delta}). 

In fact the quartic term in eq.(\ref{potential}) is what one would have 
obtained directly from 
a D-term in a supersymmetric effective field theory on the $p$-brane
world volume. In the large $N/V$ limit, the $U(1)$ gauge field coming from
the $(p+2)$-brane has a kinetic term which comes with the tension
$\frac{V}{2\pi^2}\sqrt{1+\frac{4\pi^4 N^2}{V^2}} \rightarrow N$. This means 
that the tachyon field
carries $g_1=1/\sqrt{N}$ units of charge with respect to this $U(1)$ field,
while the charge $g_2=1$ with respect to the $U(1)$ gauge field coming from
the $p$-brane. The D-term therefore gives rise to the quartic term in
eq.(\ref{potential}) where the factor $(1+\frac{1}{N})$ is just the usual
factor $(g_1^2+g_2^2)$.

We postpone the discussion of higher order terms in the
tachyon potential to section 5, where we find it easier to 
discuss it in the $T$-dual version of the present case, which will 
be analyzed in sections 3 and 4. 

Now let us discuss what happens when we have $m$ parallel $p$-branes near a
$(1,N)$ system. In this case the difference between the BPS bound for the
tension and the sum of the tensions of the individual systems is easily seen
to be:
\begin{equation}
\delta T = -\frac{TV^2}{8\pi^4\lambda} \frac{m}{N(N+m)}
\label{Nmdiff}
\end{equation}
We would now like to see if the minimization of the tachyon potential would
account for this difference. 

When the transverse distance between the $m$ $p$-branes vanishes, there is a
$U(1)\times U(m)$ gauge group, where the off-diagonal generators of $U(m)$
arise from the open strings stretched between pairs of $p$-branes. The
tachyon fields appear, as before, from the strings stretched between  the
$(p+2)$-brane and any one of the $p$-branes. There are $m$ of these tachyon
fields that transform as a fundamental representation of $U(m)$ and carry
$1/\sqrt{N}$ charge with respect to the first $U(1)$ factor. Let us label
these tachyons by $\chi^i$ for $i=1,...,m$. The scalars of the off-diagonal
gauge fields will be labelled by $A^{(ij)}_{\pm}$ and the ones along the
Cartan directions by $A^{(ii)}_{\pm}$ and $A^{(00)}_{\pm}$, where the
subscripts $\pm$ indicate the eigenvalues of the rotation operator in the
torus directions. The quartic term in the potential to the leading order in
the $V/N$ is just obtained from the supersymmetric $U(1)\times U(m)$ gauge
theory. 
Recalling the $U(1)$ charge of the tachyon, the potential involving the
tachyons and the scalar partners of the gauge fields, upto the quartic terms
is easily seen to be:
\begin{eqnarray}
{\cal{V}}&=& -\frac{V}{2\pi^3 N}\sum_{i} |\chi^i|^2 + \frac{1}{2\pi^2}
\frac{1}{N}
(\sum_i |\chi^i|^2)^2 
+\frac{1}{2\pi^2} \sum_{i}\left( \sum_{j\neq i} (|A^{(ij)}_+|^2 - 
|A^{(ji)}_+|^2)-
|\chi^i|^2\right)^2
\nonumber \\ &~&+\frac{1}{\pi^2}\sum_i \sum_{j>i} | \sum_{k} (A^{(ik)}_+
A^{(kj)}_- -
A^{(kj)}_+ A^{(ik)}_-) - {\bar{\chi}}^i\chi^j|^2  + \frac{1}{\pi^2}\sum_i
|\sum_k \chi^k A^{(ki)}_-|^2 
\label{Nmpot}
\end{eqnarray} 
If we set all the scalar partners of the gauge fields to be zero then the
tachyon potential is just the same as in the $m=1$ case considered above,
except that $|\chi|^2$ in eq.(\ref{potential}) is replaced by the $U(m)$
invariant combination $(\sum_i |\chi^i|^2)$. In any case, by using the $U(m)$
invariance, we can set all $\chi^i=0$ for $i=2,...,m$ and $\chi^1$ to be
real. Then solving for $\chi^1$ we would find the same value of the
potential as in the previous case, namely eq.(\ref{potential}), instead of
the desired value eq.(\ref{Nmdiff}). What is wrong with this ansatz of
setting all the scalar partners of
the gauge fields to zero is that, the resulting solution is unstable. Indeed
by plugging in the vev of $\chi$ in eq.(\ref{Nmpot}) for the potential, one
finds that all the scalars $A^{(1j)}_+$ for $j=2,...,m$ become tachyonic,
therefore they should also acquire vev's. Actually by using the residual
$U(m-1)$ symmetry we can set all $A^{(1j)}_+ =0$ for $j=3,...,m$ except for
$j=2$. By giving vev to $A^{(12)}_+$, one finds that the scalars
$A^{(2j)}_+$ become tachyonic for
$j=3,...,m$, necessitating a vev to one of these scalars. Clearly this will
continue. Thus many of the scalar  partners of the gauge fields also acquire
vev's.

To solve the equations, it is convenient to introduce the following notation:
\begin{eqnarray}
B^0 &=& \sum_i |\chi^i|^2 - \frac{1}{2\pi}\frac{m}{N+m} V\nonumber\\
B^k &=& \sum_{j\neq k} (|A^{(kj)}_+|^2 - |A^{(jk)}_+|^2) - |\chi^k|^2 +
\frac{1}{2\pi}\frac{1}{N+m} V \nonumber\\
X^{(ij)} &=& \sum_{k} (A^{ik}_+ A^{kj)}_- - A^{ik}_- A^{kj)}_+)-
{\bar{\chi}}^i\chi^j\nonumber\\
X^i &=& \sum_k \chi^k A^{(ki)}_-
\label{quad}
\end{eqnarray}
Note that not all the $B$'s are independent; they satisfy the relation 
$B^0 +\sum_k B^k =0$. The potential can then be rewritten as:
\begin{equation}
{\cal{V}} = -\frac{TV^2}{8\pi^4\lambda} \frac{m}{N(N+m)}  + \frac{1}{2\pi^2}
(\frac{1}{N}(B^0)^2 + \sum_k (B^k)^2)
+\frac{1}{\pi^2}\sum_i \sum_{j>i} |X^{(ij)}|^2) +\frac{1}{\pi^2}\sum_i|X^i|^2  
\label{quadv}
\end{equation}
The first term on the right hand side of the above equation is just the
right value for the BPS saturation. All the other terms above are positive
definite
and hence must be zero individually. Thus we have the equations:
\begin{equation}
B^0= B^k = X^{(ij)} = X^i = 0
\label{quadeq}
\end{equation}
The number of equations can be counted easily. $B$'s are all real, and
noting the fact that they are not all linearly independent, these give $m$
real equations, while $X^{(ij)}$'s and $X^i$'s are complex and therefore
they provide $m(m+1)$ real equations.  Thus the total number of real
equations is $m(m+2)$. 
The total number of variables (apart from the one corresponding to overall
$U(1)$) is $2m(m+1)$ real, of which $m^2$ can be set to zero, by using the
$U(1)\times U(m)$ gauge symmetry. To see this explicitly, by using the gauge
symmetry, we can set $\chi^i=0$ for $i>1$, as well as $A^{(ij)}_+ =0$ for
$j>i+1$. This still leaves the Cartan subgroup, using which we can also set
$\chi_1$ and $A^{(i,i+1)}_+$ to be real. Plugging this gauge slice in the
equations (\ref{quadeq}), one can show with some effort, that there is a
unique solution (modulo the overall $U(1)$ namely translating all the
$A^{(ii)}_{+}$ by the same amount),
namely, $\chi^1 = \sqrt{\frac{V}{2\pi}\frac{m}{N+m}}$ and
$A^{(k,k+1)}_+ = \sqrt{\frac{V}{2\pi}\frac{m-k}{N+m}}$ and all the other
fields equal to zero.

This proves, that there is a unique solution to the equations of motion,
with non-trivial vev's for the tachyon and some other massless fields, such
that the potential at this solution exactly compensates for the difference
in the tension predicted by the BPS formula for the $(m,N)$ system.

\section{$(N,1)$ Bound State for Large $V$}

Let us now consider the situation where we have $N$ $(p+2)$-branes and one of
them is carrying a $p$-brane charge due to a unit flux of magnetic field on the
torus. This is exactly the $T$ and $S$-dual of the problem considered by
Witten \cite{wit}. In our case we will be able to analyze the problem
explicitly because the instabilty and the consequent mass gap arise from
perturbative string states. BPS bound for the tension for this system is:
\begin{equation}
T(N,1) = \frac{TV}{2\pi^2 \lambda}\sqrt{N^2 + \frac{4\pi^4}{V^2}}
\label{m1tension}
\end{equation}
while for the individual systems $(1,1)$ together with $(N-1)$ of the $(1,0)$
systems the tension is 
\begin{equation}
\frac{TV}{2\pi^2\lambda}\sqrt{1+ \frac{4\pi^4}{V^2}}+ (N-1)\frac{TV}{2\pi^2
\lambda}
\label{m1tension'}
\end{equation}
The difference between these two values is small in the large volume limit and
in fact to the leading order in $1/V$ is 
\begin{equation}
-\frac{T\pi^2}{\lambda}\frac{(N-1)}{NV}
\label{m1diff}
\end{equation}
This difference, as before, should be provided by the minimization of the
potential. Upon dimensional reduction, the resulting $(p+1)$-dimensional world
volume theory will contain the usual $U(1)^N$ gauge fields $A^{(ii)}$,
$i=1,...,N$ that appear from the string going from $i$-th brane to itself.
Besides this there will be also the string states stretched between two
different branes. If the transverse distance between these branes is zero
then in the absence of magnetic flux on the first brane (say), one would obtain
this way $U(N)$ gauge fields on the $(p+2)$-brane world volume with
$A^{(ij)}$ for $i \neq j$ being the off diagonal gauge fields. Upon
dimensional reduction on the torus, the Wilson lines provide two scalars
$A^{(ij)}_{\pm}$ with the reality condition $(A^{(ij)}_+)^* =A^{(ji)}_-$. 

What happens when we turn on the unit magnetic flux on the first brane? The
magnetic field in this case is $1/V$ and once again applying the results of
\cite{call,bp,b} we find that there are tachyons in the sectors of strings 
stretched between the first and some other branes. More explicitely there is
a mass shift for the fields $A^{(1i)}_{\pm}$ and their masses are
\begin{equation}
M^2_{\pm} = (1\mp 2)\frac{2\pi}{V}
\label{m1mass}
\end{equation}
Thus $A^{(1i)}_{+}$ is tachyonic and signals an instability of the vacuum.
The corresponding Vertex operator involves twist fields that twist by an
amount $1/V$. In the large volume limit these operators go
over to the standard untwisted vertex operators for the Wilson lines.

We will again consider in this section only upto quartic term in the potential and
restrict to the leading order in $1/V$, which is expected to be
of order $((1/V)^0$. This is 
obtained by dimensional reduction of $U(N)$ gauge theory. Note that, unlike
in the previous section, the tachyon carries unit charge under the $U(1)$ of
the first brane that has the unit magnetic flux. This can be seen by looking
at the corresponding gauge kinetic term which in the large volume limit
scales to the factor $V$ just as the for the $U(1)$ gauge fields coming from
the other branes. As a result in the large volume limit, the quartic term
can be computed from an effective gauge group $U(N)$ instead of $U(1)\times
U(N-1)$. The relevant
terms in the potential can be shown to be:
\begin{eqnarray}
{\cal{V}}&=& \frac{V}{2\pi^2}\left(\frac{2\pi}{V}\sum_{i=2}^{N} 
(-|A^{(1i)}_+|^2+3 |A^{(i1)}_+|^2)
+\frac{1}{2\pi^2} \sum_i \left(\sum_{j\neq i} (|A^{(ij)}_+|^2 - 
|A^{(ji)}_+|^2)\right)^2\right .
\nonumber \\ &~&\left .+ \frac{1}{\pi^2}\sum_i \sum_{j>i} | \sum_{k\neq i,j}
(A^{(ik)}_+ A^{(kj)}_-
- A^{(kj)}_+ A^{(ik)}_-)|^2\right)   
\label{m1pot}
\end{eqnarray}

Exactly as in the previous section, we can recast the above potential in a
convenient form by defining:
\begin{eqnarray}
B^1 &=& \sum_{j\neq k} (|A^{(1j)}_+|^2 - |A^{(j1)}_+|^2) -
\frac{2\pi^3}{V}\frac{N-1}{N}  \nonumber\\
B^k &=& \sum_{j\neq k} (|A^{(kj)}_+|^2 - |A^{(jk)}_+|^2) +
\frac{2\pi^3}{V}\frac{1}{N} , ~~~~~~~~~ for ~~k>1\nonumber\\
X^{(ij)} &=& \sum_{k} (A^{ik}_+ A^{kj)}_- - A^{ik}_- A^{kj)}_+) 
\label{N1quad}
\end{eqnarray}
Note that not all the $B$'s are independent; they satisfy the relation 
$\sum_k B^k =0$. The potential can then be rewritten as:
\begin{equation}
{\cal{V}} = -\frac{\pi^2}{V} \frac{N-1}{N}  + 
\frac{V}{2\pi^2}\left(\frac{4\pi}{V}\sum_{i=2}^{N} |A^{(i1)}_+|^2 +
\frac{1}{2\pi^2} \sum_k (B^k)^2
+\frac{1}{\pi^2}\sum_i \sum_{j>i} |X^{(ij)}|^2\right ) 
\label{m1quadv}
\end{equation}
The first term on the right hand side of the above equation is just the
right value for the BPS saturation. All the other terms above are positive
definite
and hence must be zero individually. Thus we have the equations:
\begin{equation}
A^{(i1)}_+ = B^k = X^{(ij)} = 0
\label{m1quadeq}
\end{equation}
The number of equations can be counted easily. $B$'s are all real, and
noting the fact that they are not all linearly independent, these give
$(N-1)$ real equations, while $X^{(ij)}$'s and $A^{(i1)}_+$ are complex and
therefore they provide $(N+2)(N-1)$ real equations.  Thus the total number
of real equations is $N^2+2N-3$. 
The total number of variables (apart from the one corresponding to overall
$U(1)$) is $2(N^2-1)$ real, of which $(N-1)^2$ can be set to zero, by using
the $U(1)\times U(N-1)$ gauge symmetry. Thus the number of variables modulo
gauge transformation is exactly equal to the number of equations. To see
this explicitly, by using the gauge symmetry, we can set $A^{(ij)}_+ =0$ for
$j>i+1$. This still leaves the Cartan subgroup, using which we can also set
$A^{(i,i+1)}_+$ to be real. Plugging this gauge slice in the equations
(\ref{quadeq}), one can show with some effort, that there is a unique
solution (modulo the overall $U(1)$ namely translating all the
$A^{(ii)}_{+}$ by the same amount),
namely, 
$A^{(k,k+1)}_+ = \sqrt{2\pi^3\frac{N-k}{VN}}$ and all the other fields equal
to zero.

This proves, that there is a unique solution to the equations of motion such
that the potential at this solution exactly compensates for the difference
in the tension predicted by the BPS formula for the $(N,1)$. 

We can now ask the question what happens when we bring  
$m$ more $(p+2)$-branes near the above bound state. This system is exactly
the $T$ dual to the situation we have considered in Section 2. The resulting
system will
again form a $(N+m,1)$ bound state carrying $(N+m)$ units of $(p+2)$-brane
charge and 1 unit
of $p$-brane charge exactly as discussed in the previous section, but now the
expectation values are obtained from eqs.(\ref{N1quad}) and (\ref{m1quadeq}), 
with $N$ replaced by $(N+m)$. This means that the original vev's of $A^{j,j+1)}_+$
for $j=1,...,N-1$, change from $\sqrt{2\pi^3\frac{N-j}{NV}}$ to
$\sqrt{2\pi^3\frac{N+m-j}{(N+m)V}}$. On the other hand these fields are massive
fields in the $(N,1)$ bound state. Indeed,
while the imaginary parts of these fields are the Goldstone bosons, that are
eaten by the gauge fields, the real parts of these fields, as can be seen
from eq.(\ref{m1pot}) for the scalar potential, acquire a mass term through
their vev's of the form
\begin{equation}
\frac{V}{4\pi^4} \chi^i {\cal{C}}_{ij} \chi^j, ~~~ \chi^i \equiv 2
\sqrt{2\pi^3\frac{N-i}{NV}}Re(A^{(i,i+1)})
\label{chimass} 
\end{equation}
where $i,j=1,...,N-1$ and ${\cal{C}}_{ij}$ is the Cartan matrix of $SU(N)$.
Thus in the new ground state of $(N+m,1)$ bound state, some of the massive
fields intrinsic to $(N,1)$ bound state acquire vev's. We would now like to
ask the question whether one can obtain the new ground state without
discussing the details of the intrinsic massive fields of the $(N,1)$ bound
state. In other words we would
like to integrate all the massive fields and focus on the tachyonic and
massless fields. The only massless field in the $(N,1)$ bound state is the
Center of mass $U(1)$ multiplet, while in the $(m,0)$ system we have
massless $U(m)$ vector multiplet. What are the tachyonic states. At first
sight it may appear that
the fields $A^{(i,N+a)}_+$ for $i=1,...,N$ and $a=1,...,m$ are tachyonic.
But one can see by using eq.(\ref{m1pot}) for the potential, that in the
presence of the non-trivial vev's that go into the formation of $(N,1)$
bound state, only
the fields $A^{(N,N+a)}_+$ are tachyonic with mass square given by
$-\frac{2\pi}{VN}$. Furthermore these tachyon fields have cubic interactions
of the
form $-\frac{V}{2\pi^4}\chi^{N-1} \sum_{a=1}^m
(|A^{(N,N+a)}_+|^2-|A^{(N+a,N)}_+|^2$.
 
The fact that the tachyon mass square is $-2\pi/NV$ rather than $-2\pi/V$ can be
directly understood without going into the details of the intrinsic structure
of $(N,1)$ bound state. The reason is that in the process of formation of
the bound state $U(N)$ group is broken down to $U(1)/Z_N$. As a result, the
magnetic field is quantized in units of $1/NV$. The presence of the cubic
interaction has the following consequence. In the large volume limit one
would have naively expected that the order $V^0$ quartic couplings that we
are interested in 
will have a $U(m+1)$ symmetry. However if one integrates out the massive fields
$\chi^i$ then one gets an extra contribution to the quartic term
\begin{equation}
-\frac{V}{4\pi^2} {\cal{C}}^{N-1,N-1} \sum_{a=1}^m
(|A^{(N,N+a)}_+|^2-|A^{(N+a,N)}_+|^2 )^2
=-\frac{V}{4\pi^4}\frac{N-1}{N}\sum_{a=1}^m
(|A^{(N,N+a)}_+|^2-|A^{(N+a,N)}_+|^2 )^2
\label{extra}
\end{equation}
where ${\cal{C}}$ with superscripts is the inverse of the Cartan matrix and is
just the propagator of the $\chi$ fields as follows from the mass term
eq.(\ref{chimass}). Note that this term has only the $U(1)\times U(m)$ symmetry
and not the accidental $U(m+1)$ symmetry. This is because the reducible diagrams
involving the massive intermediate states involve the cubic term and the
mass term which have only the $U(1)\times U(m)$ symmetry. The net effect of
the details of the $(N,1)$ bound state is summarized in this extra term in
the quartic potential.

Now we can proceed to minimize the potential involving only the tachyon and
massless fields. As before using the $U(m)$ symmetry we can set all the fields
$A^{(N+a-1,N+b-1)}_+ = 0$ for $a=1,...,m-1$ and $b>a+1$ and set
$A^{(N+a-1,N+a)}_+ \equiv \sqrt{v^a}$ for $a=1,...,m$ to be real. 
Minimizing the potential one finds, just as before, that
$A^{(N+a-1,N+b-1)}_+ = 0$ for
$a=1,...,m-1$ and $b<a$ and $A^{(N+a-1,N+a-1)}_+ = A^{(N,N)}_+$ for
$a=1,...,m-1$. Furthermore $v^a$ satisfy the equation
\begin{equation}
{\cal{C}}_{ab} v^b - \frac{N-1}{N} \delta_{a1} v^1= \frac{2\pi^3}{NV}w_a
\label{Nmeq}
\end{equation}
where $a,b = 1,...,m$, ${\cal{C}}$ is the Cartan matrix of $SU(m+1)$ and $w_a
=\delta_{a1}$ are just the Dynkin
coefficients of the fundamental representation of $SU(m+1)$. Note that the first
term on the left hand side of the above equation comes from the usual
quartic terms that have $U(m+1)$ symmetry but the second term is due to the
additional quartic term eq.(\ref{extra}) that appears after integrating out
the massive modes. Solving this equation one finds that $v^a=2\pi^3
\frac{m-a+1}{(N+m)V}$ for $a=
1,...,m$. The minimum value of the potential is
\begin{equation}
-\frac{\pi^2}{VN}\frac{m}{N+m}
\label{Nmminpot}
\end{equation}
which is exactly the difference, to order $1/V$, between the BPS bound for the
$(N+m,1)$ system and the sum of the individual BPS bounds for $(N,1)$ and
$(m,0)$ systems. 

We can now compare the above result with the one obtained in section (2)
where we considered the bound states of $(1,N)$ system with $(0,m)$ system.  
Indeed this system is $T$-dual to the ones involving the bound state of $(N,1)$
and $(m,0)$ systems. The extra term in the quartic potential eq.(\ref{extra})
that arose by integrating out the massive modes has an exact parallel in
section (2), where it came from the fact that the charge of the tachyon with
respect to the $U(1)$ of the $(1,N)$ bound state system was $\sqrt{1/N}$.

\section{$(N,m)$ Bound States for Large $V$}

Let us now consider the bound states of $N$ $(p+2)$- branes that carry a net $m$
units of $p$-brane charge. Let $m=kN+s$, where $k$ is a non-negative integer and
$s$ is between 0 and $N-1$. If $s=0$ then we can realize this system by
giving $k$ units of flux in each $(p+2)$-brane. There are no tachyons and
the system is
just that of $N$ copies of $(1,k)$ systems whose relative separations are flat
directions. Let us then take $s$ to be between 1 and $N-1$. First consider
the case when $s$ and $N$ are relatively prime. We can realize this system
by giving
$(k+1)$ units of magnetic flux on each of the first $s$ $(p+2)$-branes and $k$
units each on the remaining $N-s$ branes. The sum of the tensions of this
system is
\begin{equation}
\frac{T}{2\pi^2\lambda} \Big( s\sqrt{V^2 +4\pi^4(k+1)^2} +
(N-s)\sqrt{V^2+4\pi^4 k^2} \Big)
\label{mNindten}
\end{equation}
On the other hand the BPS bound for a $(N,m)$ system is 
\begin{equation}
\frac{T}{2\pi^2 \lambda} \sqrt{N^2V^2 + 4\pi^4 (Nk+s)^2}
\label{mNBPS}
\end{equation}
We expect that the difference between these two values will be provided by
the minimzation of the potential. This difference in the large volume limit is
\begin{equation}
\delta{\rm T}=-\frac{T\pi^2}{\lambda}\frac{s(N-s)}{NV}.
\label{mNdif}
\end{equation}
The world volume theory of the system just described has a
$U(s)\times U(N-s)$ gauge symmetry. This is because strings stretched between 
two of the first $s$ or the remaining $N-s$ branes have identical boundary
conditions on the two ends and therefore they become massless when the relative
separations vanish, thereby giving rise to the off-diagonal gauge fields. The
tachyon fields $A^{(i,a)}_+$ for $i=1,...,s$ and $a=s+1,...,N$ are the
ground states of the open string stretched between one of the first $s$
branes and one
of the remaining $N-s$ branes. Obviously the tachyon fields transform as
fundamental representation of $U(s)$ and $U(N-s)$. Since the difference
between the magnetic fluxes at the two ends of the string are 1 unit each,
it follows from  \cite{call,bp,b}, that the multiplicity of these
tachyons is one and their masses squares are $-2\pi /V$ each. 
 
The quartic term in the potential involving massless and
tachyonic fields to order $(\frac{1}{V})^0$, is obtained just by dimensional
reduction from $(p+2)$ to $p$-brane world-volume. The relevant terms in the
potential then are:
\begin{eqnarray}
{\cal{V}}&=& \frac{V}{2\pi^2}\left(\frac{2\pi}{V}\sum_{i=1}^{s}\sum_{a=s+1}^N
(-|A^{(ia)}_+|^2 +3
|A^{(ai)}_+|^2)
+\frac{1}{2\pi^2} \sum_i \left(\sum_{j\neq i} (|A^{(ij)}_+|^2 - |A^{(ji)}_+|^2)
\right)^2 \right.
\nonumber \\ &~&+\left.\frac{1}{\pi^2}\sum_i \sum_{j>i} | \sum_{k\neq i,j}
(A^{(ik)}_+ A^{(kj)}_-
- A^{(kj)}_+ A^{(ik)}_-)|^2\right) .   
\label{mNpot}
\end{eqnarray}
Here the indices $i,j,k$ run from 1 to $N$ unless stated otherwise. We can
again recast this potential into a convenient form by defining:
\begin{eqnarray}
B^k &=& \sum_{j\neq k} (|A^{(kj)}_+|^2 - |A^{(jk)}_+|^2) -
\frac{2\pi^3}{V}\frac{N-s}{N} , ~~~~~~~~~ {\rm for} ~~1\leq k \leq s \nonumber\\
B^k &=& \sum_{j\neq k} (|A^{(kj)}_+|^2 - |A^{(jk)}_+|^2) +
\frac{2\pi^3}{V}\frac{s}{N} , ~~~~~~~~~{\rm for} ~~s<k\leq N\nonumber\\
X^{(ij)} &=& \sum_{k} (A^{(ik)}_+ A^{(kj)}_- - A^{(ik)}_- A^{(kj)}_+) 
\label{m1quad}
\end{eqnarray}
As before, not all the $B$'s are independent; they satisfy the relation 
$\sum_k B^k =0$. The potential can then be rewritten as:
\begin{equation}
{\cal{V}} = -\frac{\pi^2}{V} \frac{s(N-s)}{N}  + 
\frac{V}{2\pi^2}\left(\frac{4\pi}{V}\sum_{i=1}^{s}\sum_{j=s+1}^N |A^{(ji)}_+|^2 +
\frac{1}{2\pi^2} \sum_k (B^k)^2
+\frac{1}{\pi^2}\sum_i \sum_{j>i} |X^{(ij)}|^2\right). 
\label{m1quadv1}
\end{equation}
The first term on the right hand side of the above equation is just the
right value for the BPS saturation. All the other terms above are positive
definite
and hence must be zero individually.  
The number of equations can be counted easily. $B$'s are all real, and
noting the fact that they are not all linearly independent, these give
$(N-1)$ real equations. $X^{(ij)}$'s are complex and therefore they provide
$N(N-1)$ real equations.  Finally $A^{(ji)}_+ =0$ for $i=1,...,s$ and
$j=s+1,...,N$ give $2s(N-s)$ real equations. Thus the total number of real
equations is $(N^2+2Ns-2s^2-1)$. 
The total number of variables (apart from the one corresponding to overall
$U(1)$) is $2(N^2-1)$ real, of which $(s^2 + (N-s)^2-1)$ can be set to zero,
by using the $U(s)\times U(N-s)$ gauge symmetry. Thus the number of
variables modulo gauge transformation is exactly equal to the number of
equations. However, in this case, since the gauge group is smaller, there
could be several gauge inequivalent solutions. We have not analyzed this
problem in detail, but in the following we show that there always exists at
least one solution, which is compatible with the BPS bound.  

Let us take the ansatz $A^{(ij)}_+=0$ for
$j\neq i,i+1$,
and $A^{(ii)}_+=A^{(11)}_+$ and $A^{(i,i+1)}_+ \equiv \sqrt{v^i}$ are real.
$X^{(ij)}$'s then are identically zero. From the expression for $B^i$'s it
follows, that $v^i$'s for $i=1,...,N-1$ must satisfy:
\begin{equation}
\sum_{j=1}^{N-1} {\cal{C}}_{ij}v^j = \frac{2\pi^3}{V} \delta_{is} \equiv
\frac{2\pi^3}{V} w_i
\label{mNeq}
\end{equation}
where ${\cal{C}}$ is the Cartan matrix of $SU(N)$. It is clear that the $w_i$'s
are just the Dynkin coefficients of the $s$-th rank
anti-symmetric tensor representation of $SU(N)$. Thus if $\alpha_i$ are the
simple roots of $SU(N)$ labelling the $i$-th point in the Dynkin diagram,
then $\sum_i \alpha_i v^i$ is $2\pi^3/V$ times the highest weight of the $s$-th
rank anti-symmetric tensor representation. Explicitly the solution is
$v^r=2\pi^3 r(N-s)/NV$ for $r \leq s$ and for $r>s$, $v^r= 2\pi^3
s(N-r)/NV$. The minimum
value of the potential is 
\begin{equation}
{\cal{V}}_{min} = -\frac{1}{\pi}v^s + \frac{V}{4\pi^4} \sum_{i,j=1}^{N-1}
v^i v^j 
{\cal{C}}_{ij} = -\pi^2\frac{s(N-s)}{NV}
\label{mNmin}
\end{equation}
This is exactly the desired difference  eq.(\ref{mNdif}).  We do not know whether 
the above solution is unique, but
we believe that for $m$ and $N$ coprime, it is so modulo the gauge
transformations.

In the above discussion, actually we have not really used the fact that $s$ and
$N$ are relatively prime. In fact the solution described above also seems to
hold for the case when $s$ and $N$ are not relatively prime. The minimum of
the potential of course gives the correct BPS bound for this situation also.

Let us consider now the case where $s$ and $N$ have gratest common divisor $c$,
that is $s=r.c$ and $N=n.c$, for some positive integers $r$ and $n$ that are
relatively prime, with  $r<n$, since $s<N$. Then this system is
equivalent to $c$ identical copies of
$(n,kn+r)$ bound states which are each BPS states. Therefore, there will be no
mutual force between these $c$ copies. 
We will now show, that there exist other solutions giving the correct
minimum. To exhibit this, consider the ansatz
\begin{eqnarray}
A^{(ij)}_+ &=& 0, ~~~~~~~~~~ j \neq i, i+c, \nonumber \\
A^{(ii)}_+ &=&A^{(11)}_+, \nonumber \\
A^{(i,i+c)}_+ &=& real. 
\label{ansatzmN}
\end{eqnarray}
Denoting $v^i=(A^{(i,i+c)}_+)^2$, we can see that the equations of motion 
arising from the potential (\ref{mNpot}) are consistent with this ansatz
provided $v$'s satisfy the equation:
\begin{equation}
\sum_{b=1}^{n-1} {\cal{C}}_{ab}v^{i+(b-1)c} = \frac{2\pi^3}{V}\delta_{ar}\equiv
\frac{2\pi^3}{V}w_a, ~~~~~~~ i=1,...,c
\label{mNequation}
\end{equation}
where ${\cal{C}}$ is the Cartan matrix of $SU(n)$. Thus we have $c$ sets of
uncoupled equations, each involving $SU(n)$ Cartan matrices. $w$'s are just
the Dynkin coefficients of the $r$-th rank anti-symmetric tensor
representation of $SU(n)$. It is clear from the previous analysis that the
potential at the minimum is
\begin{equation}
-c \pi^2 \frac{r(n-r)}{nV}
\label{divmin}
\end{equation}  
This is exactly the difference expected from eq.(\ref{mNdif}).
 
In the next section we will discuss the surviving gauge group
beyond the quartic approximation we used so far, finding that,
indeed, in the coprime case it is just the center of mass $U(1)$,
whereas in the noncoprime case it is of rank $c$.

.

\section{Higher order terms and Torons}

In the last sections, we restricted the analysis to the quartic potential
for the tachyonic fields. In the Appendix we show how this potential
arises from a tree level string computation involving four external tachyon
vertex operators, after integrating out the massive Kaluza-Klein modes 
$\phi(n_1,n_2)$ corresponding to the Cartan directions. These are the
only massive modes which are relevant at the quartic level.

In this section we address the question of higher order
contributions to the tachyon potential. We will discuss this in the context
of the examples presented in sections 3 and 4 in which the relevant 
perturbation parameter is $1/V$. We will use the effective field theory
on the world-volume of the $(p+2)$-brane and will carry out a
Kaluza-Klein analysis in the presence of a magnetic flux on the
two-torus part of the world-volume.
 
The higher order contributions to the effective potential  can come
either from higher derivative terms in $(p+2)$ dimensions (like $F^4$)
or from the reducible diagrams of higher point functions. It is easy
to see that the former contributions are suppressed in the large
volume limit. However, the reducible diagrams involving vertices 
due to $F^2$ term give rise to contributions which are of the same order 
in volume as the quartic term considered previously. 
In fact it is easy to see this by noting that the n-point vertices scale
as $V^{\frac{n-4}{2}}$ and the the massive propagators scale as $V$. It
then follows that the term $|\chi|^{2n}$ scales as $V^{n-2}$. The
above massive modes include, in addition to the $\phi$'s, also the
higher Landau levels. The fact that these massive modes contribute
to the higher order terms in the tachyon potential implies that
they also acquire expectation values. A more convenient way of 
including these infinite sets of modes, when minimizing the potential,
is to work directly with the $(p+2)$ dimensional field strength.

As a simple example, let us discuss the $(2,1)$ system  considered
as a bound state of a $(p+2)$-brane which has one unit of magnetic flux
on it with another $(p+2)$-brane. The world-volume theory of the $(p+2)$ 
-branes involves a $U(2)$ theory, where the $U(1)$ corresponding to
the first brane has one unit of magnetic flux. If $(x_1,x_2)$ are the
compact world-brane coordinates of the $(p+2)$-brane, corresponding
to a torus $T^2$ with volume $V=R_1R_2$, the gauge potential can
be taken to be $A^{(1,1)}_1=0$ and $A^{(1,1)}_2=2\pi x_1/V$, with field strength
$2\pi/V$. 
Writing $U(2)=SU(2)\times U(1)/Z_2$, where $U(1)$ corresponds to the center of
mass degree of freedom, we can decompose the background gauge potential as
$A_z=-\frac{i\pi x_1}{2V}(I+\sigma_3)$, where $z=x_1+ix_2$ and $I$ is the
identity matrix,
one can see that the field strength of the center of mass $U(1)$
has half unit of flux. The contribution to the energy of the center of mass $U(1)$
due to the half unit flux is $\pi^2/2V^2$ and saturates the BPS bound.
This implies that contribution of the $SU(2)$ part to the ground
state energy should vanish. In order to see this, let us expand the  
$SU(2)$ gauge potential around the background as follows:
\begin{equation}
A_z=A^0_z+\phi\sigma_3+\xi\sigma_++\bar{\eta}\sigma_-,~~~~~~A_{\bar{z}}=
(A_z)^\dagger.
\label{fluct}
\end{equation} 
where $\sigma_{\pm}=\sigma_1\pm i\sigma_2$ 
and $A^0_z=-i\pi x_1\sigma_3/2V$. Note that $A^0_z$ obeys the
following boundary conditions:
\begin{eqnarray}
A^0(x_1,x_2+R_2)&=&\Omega_2A^0(x_1,x_2),\nonumber\\
A^0(x_1+R_1,x_2)&=&\Omega_1A^0(x_1,x_2),
\label{bc}
\end{eqnarray}
where $\Omega_2=1$, $\Omega_1=\exp(i\pi x_2 \sigma_3/R_2)$ and
$\Omega A$ means the gauge transformation of $A$ by $\Omega$.
These are in fact the boundary conditions of an $SU(2)$ toron:
\begin{equation}
\Omega_2(R_1)\Omega_1(0)=-\Omega_1(R_2)\Omega_2(0).
\label{toron}
\end{equation}
The boundary conditions for the fluctuations $\phi$, $\xi$ and
$\eta$ are dictated by the requirement that $A$ satisfies the
same boundary conditions as $A^0$. This implies that the fluctuations are 
periodic under $x_2\rightarrow x_2+R_2$ while under $x_1\rightarrow
x_1+R_1$ they transform as:
\begin{equation}
\phi\rightarrow \phi,~~~~~\xi\rightarrow e^{2\pi i x_2/R_2}\xi,~~~~~
\eta\rightarrow e^{2\pi i x_2/R_2}\eta.
\label{bcchi}
\end{equation}
The $F^2$ term corresponding to  $U(2)$ is the sum of the
ones corresponding to the center of mass $U(1)$ and the relative
$SU(2)$.  The
$SU(2)$ contribution is of the form $\frac{1}{2}\int_{T^2}(F_3^2+
|F_+|^2)$, where:
\begin{eqnarray}
F_3&=&\frac{\pi}{V}-i(\partial_{\bar{z}}\phi-\partial_z\bar{\phi})
+|\eta|^2-|\xi|^2\nonumber\\ 
F_+&=&D_z\xi+D_z^{\dagger}\eta+2i\bar{\phi}\xi-2i\phi\eta.
\label{fs}
\end{eqnarray}
Here $D_z=\partial_z-i\pi x_1/V$ and $D_z^\dagger=-\partial_{\bar z}
+i\pi x_1/V$. Since the $SU(2)$ contribution 
is the sum of two integrals of
semi-positive definite functions on $T^2$, the minimum energy
configuration is obtained when $F_3=0$ and $F_+=0$.
We can recover, in this framework, the result on the quartic 
tachyon potential discussed in the appendix if in the term
$\int_{T^2} F_3^2$ we restrict ourselves to the lowest Landau
level (i.e., $D_z\xi=0$) and integrate out the massive Kaluza-Klein
modes of $\phi$. The lowest Landau level is explicitly given by
\begin{equation}
\Psi_0=(-2\pi i\tau)^{1/4} e^{-\pi x_1^2 /V} \theta_3(\tau|{\bar z});~~~~~
\tau\equiv i\frac{R_1}{R_2},
\label{ground}
\end{equation}
where $\theta_3 $ is a Jacobi theta function. Note that since $\phi$
is neutral its Kaluza-Klein expansion is just the usual Fourier mode
expansion whereas $\xi$ and $\eta$ are expanded in the Landau modes
given by $(D^\dagger_z)^n\Psi_0$. However as an inspection of the 
equation (\ref{fs}) reveals, due to the fact that the massive
modes of $\phi$ acquire non-trivial expectation value, the higher
Landau modes can not be ignored. 

The most convenient way of
including these higher modes is to look for the solutions to the
equations $F_3=F_+=0$ on $T^2$.     
As shown
in \cite{Amb, thooft}, there exist such zero field strenght 
configurations with
non-zero $Z_2$ flux arising from the twisted boundary conditions
(\ref{toron}). (For a discussion of torons in the context of D-branes
see \cite{Ram}.) 
 For example, by making a suitable gauge transformation we
can map the boundary conditions (\ref{bc}) to the
the constant ones $\Omega_1=i\sigma_1$ and $\Omega_2=
i\sigma_3$. It is clear that zero gauge potential
satisfies the latter boundary conditions, and therefore
a gauge potential obeying (\ref{bc}) is a 
pure gauge.

In the above gauge in which $\Omega_1=i\sigma_1$, $\Omega_2=
i\sigma_3$ it is easy to see that the relative $SU(2)$ is broken.
The kinetic term for the gauge potential $A_\mu$, where $\mu$ refers
to the $p+1$ noncompact directions, comes from $|F_{\mu z}|^2$. Then
the massless gauge fields are indepedent of the $T^2$ coordinates
$z,\bar z$. Since $A_\mu$ satisfies the boundary condition (\ref{bc})
this implies that it should commute with $\sigma_1$ and $\sigma_3$.
It follows that it can only be the identity matrix, that is, only the
center of mass $U(1)$ is unbroken.  

This argument can be extended to the case of $N$ number of $(p+2)$-branes 
with $m$ units of magnetic flux distributed among them, as discussed in
section 4. Following the same steps as above, one is lead to consider
the $SU(N)$ toron configuration corresponding to a $Z_N$-twist given by
\begin{equation}
\Omega_2(R_1)\Omega_1(0)=\Omega_1(R_2)\Omega_2(0)e^{2i\pi m/N}.
\label{toronN}
\end{equation}
Following 't Hooft, we can write constant $\Omega$'s satisfying
this condition in terms of the matrices $P$ and $Q$ introduced
in \cite{thooft} as:
\begin{equation}
\Omega_1=P^m,~~~~~~~~\Omega_2=Q.
\label{omegas}
\end{equation}
Then the question of the unbroken gauge symmetry reduces to finding
constant (i.e., independent of $z$ and $\bar z$) $SU(N)$ matrices $A$ 
which commute with $P^m$ and $Q$. Since $Q$ is diagonal then $A$ must
be diagonal. To analyse the conditions coming from  $[A,P^m]=0$,
one has to distinguish two cases: (i) $N$ and $m$ being coprime and (ii) 
not coprime, 
i.e., $N=cn$ and $m=cs$ for coprime integers $n$ and $s$. In case (i) the
$SU(N)$ is completely broken whereas in (ii) there is an unbroken rank
$c-1$ gauge group, in addition to the center of mass $U(1)$. 
 
The last point we would like to make concerns the spectrum of the massive
K-K modes. We would like to show that the massive spectrum in the toron background is indeed the correct one.
For the $(1,N)$ system ( the example considered in section 2),
let us consider the spectrum of the massive K-K modes of the open string 
from the $(p+2)$-brane to itself. Due to the presence of $N$ units of 
magnetic flux, the momenta $p_1$ and $p_2$ are rescaled \cite{call} by a factor
$1/\sqrt{1+(\frac{2\pi^2N}{V})^2}$ which in the small volume limit
becomes $V/2\pi^2N$. This implies that the momenta are given by 
$p_i=\frac{R_i}{\pi}(n_i+\frac{k_i}{N})$ for $k_i=0,1,\dots ,N-1$ and arbitrary 
integer $n_i$.     
On the other hand, in the dual $(N,1)$ system due to $Z_N$ twistings along
the two one-cycles of the torus the K-K spectrum in
the background of the $SU(N)$ toron is given by momenta that lie in a shifted lattice.
Since the boundary conditions $\Omega_1=P$ and $\Omega_2=Q$ commute upto the center of the group, it is clear that they can be simultaneously diagonalized in the adjoint representation.
Thus we can find a basis of $U(N)$ generators such that $\Omega_i^{-1} J_{(k_1,k_2)} 
\Omega_i = e^{i2\pi \frac{k_i}{N}} J_{(k_1,k_2)}$. With a little bit of effort one can show that there is a one to one correspondence between the $N^2$ generators of $U(N)$ and the twists
$(k_1,k_2)$ with $k_i=0,1,\dots, N-1$. Thus the K-K momenta in the toron background  are
$\frac{2\pi}{R_i}(n_i+\frac{k_i}{N})$, where $n_i$ and $k_i$ are as before. 
Thus the two spectra agree, if one takes into account the $T$-duality 
$R_i\rightarrow 2\pi^2/R_i$. 

\section{Conclusions}

In this paper we have studied the formation of bound states between
$p$ and $(p+2)$-branes. One of the simplifying feature of this analysis
was that the dynamical mechanism responsible for the bound state can
already be studied at the string tree level. In fact the binding energy
should appear entirely at the tree level, as it scales like inverse power of
string coupling. We have argued here that the tachyons that appear in the
open strings stretched between $p$ and $(p+2)$-branes stabilize the
potential. In general, however, determination of the minimum of the
potential requires the exact tree level potential. In the present work
we have considered limiting cases of large or small volume. In sections
2,3 and 4 we determined the binding energy from the potential containing
up to quartic order terms in the tachyon fields and showed that it agrees
with the prediction of the BPS formula. As we discussed in section 5,
to the leading order in $1/V$  also terms of higher order in the
tachyon potential do contribute, after integrating out the massive modes.
This is because these massive modes are becoming massless in the large 
$V$ limit. Or in other words, in the renormalisation group language, 
irrelevant operators are becoming marginal in that limit. 
Of course, as a result of this the expectation value of the 
tachyon fields are expected to be modified by higher order terms. 
However, as we saw in section 5, the value of the potential at the minimum 
does not change. Moreover, in section 5 we discussed the surviving gauge
group at the minimum and showed that for $(N,m)$ system it is $U(1)$
in the coprime case and it is $(U(1))^c$ in the non-coprime case, $c$ being the
greatest common divisor of $N$ and $m$.
 
One can ask the question, what happens in the finite volume case. In this
case, one needs the exact tree level potential involving tachyonic and
massless fields. However, the string dualities (in particular the $SL(2,Z)$ of
type IIB) predicts the existence of the bound states. This means that in the
sigma-model framework, if one turns on the relevant boundary operators
corresponding to the tachyons, there should exist a non-trivial fixed point,
and the so-called $g$-function (namely the tree level partition function)
should be given exactly as predicted by the BPS bound.
Thus type IIB $SL(2,Z)$ symmetry predicts non-trivial fixed points in a
whole variety of bounadry conformal field theories that are perturbed by
relevant boundary operators. Let us consider the case of the bound state
between $(1,N)$ and $(0,1)$ systems discussed in section 2. Since the
operators in question are twist operators that represent string stretched
between two different branes
(say between a $(p+2)$ and a $p$-brane,  the anti-twist operator being the
one that reverses the orientation) it is clear that in an arbitrary
correlation function involving $n$ twist and $n$ anti-twist operators, the
operators are ordered, ie. two adjacent operators on the boundary of the
disc must be twist and anti-twist respectively; they cannot be of the same
type. This is exactly the kind of situation one encounters in the spin-1/2
Kondo problem \cite{kondo},  
however the precise operators appearing in the Kondo problem
involves Sine-Gordon type fields whereas in our case we have complicated
twist fields. What is common between the two problems is the appearance of
cocycles that are in the spin-1/2 representation, and in this sense we can
refer to them as some generalization of the Kondo problem. 

In the more general example discussed at the end of section 2 (namely the
bound state of $(1,N)$ and $(0,m)$ systems) as well as in sections 3 and 4,
we have several branes and minimization of the potential involves giving
vev's simultaneously to strings stretched between different pairs of branes.
In all of these cases there is an underlying $A_n$ type Dynkin diagram. The
operator
$A^{(i,i+1)}_+$ takes one from the $i$-th brane to $(i+1)$-th one and its
complex conjugate takes one the opposite way. The number of these operators
is equal to the number of simple roots. The cocycles $S^i_{\pm}$, coming
with these operators, for $i$ running over the simple roots, satisfy the
condition
\begin{eqnarray}
S^i_+.S^j_+ &=& 0,~~~~~~for j \neq i+1 \nonumber \\
S^i_-.S^j_- &=& 0,~~~~~~for j \neq i-1 \nonumber \\
S^i_+.S^j_- &=& 0,~~~~~~for j \neq i
\label{dynkin}
\end{eqnarray}
Thus $S^i_{\pm}$ are just the generators corresponding to the simple roots
(and their negatives respectively) in the fundamental representation of
$A_n$ algebra. In any given correlation function, one has to take the trace
over the products of these cocycles, appearing in the order the operators are
inserted on the boundary of the disc. The
operators that multiply these cocycles involve either the twisted or
untwisted operators depending on whether they are relevant or marginal (not
truly marginal) operators. The cases we have considered in this paper are
the ones where the relevant operators in question are nearly marginal. It is
an interesting open question whether such
boundary perturbations, when the relevant operators are far from being
marginal, admit non-trivial fixed points and if so whether the
disc partition functions (ie. the $g$-functions) at the new fixed points are
in accordance with the values predicted by the string duality.
The fact that there is a tachyon instability and that the BPS bound implies
that the potential is bounded below, suggests strongly that such non-trivial
fixed points should exist.

There is another direction in which one can apply the results of this paper,
namely the systems involving $p$, $(p+2)$ and $(p+4)$-branes \cite{Ram}. Again the
existence of $(p+2)$ branes, break the supersymmetry and give rise to tachyons,
whose vev's should restore the supersymmetry and saturate the BPS bounds.

Finally, one can ask what our analysis would imply on the bound
states between a D-3-brane and a fundamental string. Indeed this is just
$S$-dual of the present discussion. Under $S$-duality, a D-3-brane with
magnetic flux on a torus, becomes a D-3-brane with electric field in the
non-compact 1+1 world volume. On the other hand D-string becomes an
infinitely long fundamental string. The tachyon which was an elementary open
string stretched between the D-3-brane and D-string, and carried electric
charge with respect to the $U(1)$ of the D-3-brane, now becomes an open
D-string stretched between the D-3-brane and the infinitely long NS-string,
carrying a magnetic charge with respect to the $U(1)$ of the D-3-brane.
Thus, in the world volume $U(1)$ theory of the D-3-brane, there should
appear tachyonic magnetic monopoles (solitonic states), as one brings the
fundamental string close to it. This situation is similar to the case
studied by Polchinski and Strominger \cite{PS} in connection with breaking
the $N=2$ 4-d space-time supersymmetry by turning on field strengths in type
IIA compactification on a Calabi-Yau space. In their discussion, the theory
becomes tachyonic near the conifold singularity (with the would be massless
soliton becoming tachyon), and stabilizing the potential with respect to
this tachyon field, restores the supersymmetry. Their discussion can also be
extended to the $N=4$ case (by considering type IIA on $K_3 \times T^2$ with
non-trivial field strengths) and a similar phenomenon happens when one
shrinks appropriate 2-cycles of $K_3$. The would-be massless solitonic
states, that enhance some $U(1)$ to $SU(2)$, becomes tachyonic whose vev
again restores the supersymmetry.

{\bf{Acknowledgements}} 

We would like to thank T. Jayaraman for collaboration in the initial stages
of this work. We also thank I. Antoniadis, G. Mussardo and G. Thompson for 
very useful discussions. One of us (K.S.N.) would like to thank 
K. Gopalakrishnan for computer
facilities while in India, where part of this work was done.

\begin{flushleft}
{\large\bf Appendix} \end{flushleft}
\renewcommand{\theequation}{A.\arabic{equation}}
\renewcommand{\thesection}{A.}
\setcounter{equation}{0}

In this Appendix, we will compute the 4-pt. function of tachyons, that was
required for the determination of the quartic potential in Section 2. 
Tachyon is the ground state of the open string stretched between the
$(p+2)$-brane (with large magnetic field $2\pi N/V$) and the $p$-brane. As
described in Section 2, the string modes in this sector are not integer
moded. In fact they are shifted by $\epsilon = \frac{1}{\pi} \arctan 
(V/2\pi^2 N) = V/2\pi^3 N$ to the
leading order in $V/N$ If $X^2$ and $X^3$ are the coordinates on the torus,
then the complex combination $X=X^2+iX^3$ is twisted by $\epsilon$. Thus the
corresponding vertex operators involve the twist fields (tigether with the
twist fields for the fermionic partners). Let us denote by $\sigma_+$ the
bosonic twist field that generates this boundary condition. Then $\sigma$
takes one from the $p$-brane to the $(p+2)$-brane and $\sigma_-$ (the
anti-twist field) the opposite way. We are then interested in the
correlation function of 2 $\sigma_+$'s and 2 $\sigma_-$'s,
on the boundary of a disc (or equivalently the upper half plane, which is
what we shall be using here) with an ordering such that 2 twist fields (or
the 2 anti-twist fields) are never adjacent to each other. Of course we can
fix three of the positions using the $SL(2,R)$ invariance of the upper-half
plane. Let the positions of the two twist fields be $x_1 = 0$ and $x_3= 1$
and the two anti-twist fields be $x_2=x$ and $x_4 = \infty$, where $x$ is
real and lies between 0 and 1. This means that the the boundary conditions
on the interval
$(-\infty,0)$ and $(x,1)$ are Dirichlet, and on the other two intervals, the
twist (or anti-twist fields) ensure the right boundary condition of Neumann
shifted by the magnetic field. In the large magnetic field limit, the latter
go over to Dirichlet condition. On the Dirichlet intervals $(-\infty,0)$ and
$(x,1)$, the value of $x$ must be the position of the $p$-brane (say $X_0$). 
However, going from the first to the second Dirichlet interval, the string
can wind around various cycles of the torus, which would give the
world-sheet instanton contributions. First let us consider the quantum part
of the correlation function, namely with winding number zero sector.

Following the work of Dixon et al \cite{dixon}, we can compute the normalized
correlator  
\begin{equation}
\tilde{G}(z,w,x_i) = \frac{\langle \partial_z X(z,\bar{z}) \partial_w
\bar{X}(w,\bar{w})
\sigma_+(x_1) \sigma_-(x_2) \sigma_+(x_3) \sigma_-(x_4)\rangle}
{\langle \sigma_+(x_1) \sigma_-(x_2) \sigma_+(x_3) \sigma_-(x_4)\rangle}
\label{G}
\end{equation}
Then $\tilde{G}$ has the following form:
\begin{eqnarray}
\tilde{G}= -\omega(z) {\tilde{\omega}} (w) &[&
\frac{\epsilon (z-x_1)(z-x_3)(w-x_2)(w-x_4)}
{(z-w)^2} \nonumber \\ &+& 
\frac{(1-\epsilon)(z-x_2)(z-x_4)(w-x_1)(w-x_3)}{(z-w)^2} + 
A(x_i)] \label{Gansatz}
\end{eqnarray}
where 
\begin{equation}
\omega(z) = (z-x_1)^{\epsilon}(z-x_2)^{(1-\epsilon)}(z-x_3)^{\epsilon}(z-x_4)
^{(1-\epsilon)}
\label{omega}
\end{equation}
and $\tilde{\omega}(z)$ is the same as $\omega$ but $\epsilon$ replaced by
$(1-\epsilon)$. $A(x_i)$ is an arbitrary function of $x_i$'s subject to the
boundary condition that the winding number is zero, namely $\int_a^b dX =0$,
wher $a$ and $b$ are respectively points on the two Dirichlet intervals.
This can be made more explicit by considering another normalized correlator:
\begin{equation}
\tilde{H}(\bar{z},w,x_i) = \frac{\langle \partial_{\bar{z}} X(z,\bar{z})
\partial_w \bar{X}(w,\bar{w})
\sigma_+(x_1) \sigma_-(x_2) \sigma_+(x_3) \sigma_-(x_4)\rangle}
{\langle \sigma_+(x_1) \sigma_-(x_2) \sigma_+(x_3) \sigma_-(x_4)\rangle}
\label{H}
\end{equation} 
The Dirichlet boundary condition on the two intervals, implies that
$\tilde{H}(\bar{z},w,x_i)=-\tilde{G}(\bar{z},w,x_i)$, as one approaches the
cuts along the two Dirichlet intervals. Note that with this condition, the
boundary conditions on the remaining two intervals corresponding to
$(p+2)$-branes with magnetic flux, namely $e^{i\pi\epsilon} \tilde{G} =
e^{-i\pi\epsilon}\tilde{H}$, are automatically ensured due to the cuts in 
$\omega$. The
zero instanton sector, namely $\int_a^b (dz \tilde{G} +d\bar{z}\tilde{H})=0$,
implies (taking $w\rightarrow \infty$ and setting the values of $x_i$ using
$SL(2,R)$  mentioned above) that
\begin{equation}
A(x) = \frac{\int_0^x (z-x)\omega(z)}{F(x)} = x(1-x) \partial_x
\log F(x)
\label{A}
\end{equation}
where $F(x)$ is the hypergeometric function
\begin{equation}
F(x) = \int_0^x \omega(z) =\int_0^1 dx z^{-\epsilon}
(1-z)^{-1+\epsilon}(1-zx)^{-\epsilon}
\label{hyper}
\end{equation}

Recalling the OPE of the stress energy tensor $T(z)= -:\partial_z X
\partial_z \bar{X}:$ with the primary field
$\sigma_-(x)$, we can deduce a differential equation with respect to $x$
for the correlation function $G(x)=\langle \sigma_+(0) \sigma_-(x)
\sigma_+(1) \sigma_-(\infty)\rangle$ (after fixing the $x_i$'s as
mentioned above), and using the eq.(\ref{A}), we find, exactly as in
ref.[\cite{dixon}]:
\begin{equation}
G(x) = [x(1-x)]^{-2\Delta} \frac{1}{F(x)}, ~~~~~~~~ \Delta =
\frac{1}{2}\epsilon(1-\epsilon)
\label{Gdelta}
\end{equation}
where $\Delta$ is the dimension of the twist field. 

Now we can include also the instatnton sectors. For a winding sector
labelled by the lattice vector $L= n_1 R_1 +in_2 R_2$ for $X$ between the
two Dirichlet intervals, we have $X(z,\bar{z}) = \frac{L}{F(x)}(\int_a^P
dz\omega + \int_a^P d\bar{z} \bar{\omega})$ and similarly
$\bar{X}(z,\bar{z}) = \frac{\bar{L}}{F(x)}(\int_a^P dz\omega' + \int_a^P
d\bar{z} \bar{\omega}')$. One can then show, that the instanton action
contributes to the correlation function multiplicatively, by the exponential
of the classical action, which is $e^{-\tau(n_1^2 R_1^2 + n_2^2 R_2^2)}$, where 
\begin{equation}
\tau = \frac{F(1-x)}{2\pi^2\sin(\epsilon) F(x)}
\label{t}
\end{equation}
is positive and real for real $x$ between 0 and 1.

The complete tachyon vertex operator in $(-1)$-ghost picture is $e^{-\phi}
e^{ip_{\mu} X^{\mu}} \sigma_{\pm} e^{\pm i(1-\epsilon)\phi_1}$, where $\phi$
is the bosonization of the superghost and $\phi_1$ is the bosonization of
the fermioninic partners of $X$ and $\bar{X}$. Finally, $p_{\mu}$ is the
$(p+1)$ dimensional world-volume momenta, satisfying the onshell condition
$\sum_{\mu} p_{\mu}^2 = m^2 = \epsilon$. Since the total ghost charge of
$\phi$ on the disc must be $-2$, we must insert two operators in $(-1)$
ghost picture and two in 0-ghost picture. The latter are obtained by
applying the picture changing operators on the $(-1)$-picture vertex operator.
Taking both the twist fields in $(-1)$-ghost picture, we see that only the
world-volume part of the picture changing operator contributes, with the
final result for the 4-pt function $I(x)$ involving the 2-tachyons and their
complex conjugates:
\begin{equation}
I(x) = p_1.p_3 x^{s-1} (1-x)^{t-1} \frac{1}{F(x)} \sum_{n_1,n_2} 
e^{-\tau(n_1^2 R_1^2 + n_2^2 R_2^2)}
\label{I}
\end{equation}
where we have also included all the instanton sectors. In the above expression,
$s$ and $t$ are the Mandelstam variables $(p_1+p_2)^2$ and $(p_2 + p_3)^2$
respectively, where $p_i$ are the on-shell world-volume momentum carried by
the operators at $x_i$.

Few comments are in order. As $x\rightarrow 0$, $F(x)\rightarrow 1$ while 
$F(1-x) \rightarrow sin(\epsilon)(-\log (x) + \log(\delta))$ where $\log(\delta)
=1/\epsilon$ in the limit $\epsilon \rightarrow 0$,
which is the case of interest to us here. Thus in $x\rightarrow 0$ limit,
$I(x) \rightarrow p_1.p_3 x^{s-1}(\frac{x}{\delta})^{(n_1^2 R_1^2 + n_2^2
R_2^2)/2\pi^2}$, which just means that the fields $\sigma_+$ and $\sigma_-$ 
go to an intermediate open string state from the $p$-brane to itself 
and carrying the winding numbers $n_1$ and $n_2$.
Indeed in the sector of the open string from the $p$-brane to itself, there are
precisely such winding states. $C(n_1,n_2) \equiv \delta^{-\frac{1}{4\pi^2}(n_1^2 
R_1^2 + n_2^2 R_2^2)}$ (modulo some phases) are just
the structure constants for this 3-pt function. Integrating $x$ near $0$ 
we find a contribution of the following form:
\begin{equation}
\sum_{n_1,n_2}\frac{p_1.p_3}{s+m^2}|C(n_1,n_2)|^2 ~~~~~~~~
m^2 = (n_1^2 R_1^2 +n_2^2 R_2^2)/2\pi^2   
\end{equation}
This result can be understood in terms of factorized diagrams arising 
from the following three point functions:
\begin{equation}
\sum_{n_1,n_2}(C(n_1,n_2) \bar{\chi}\partial_{\mu} \chi 
A_{\mu}(n_1,n_2)+c.c.) 
+\sum_{(n_1,n_2)\neq (0,0)}(m C(n_1,n_2) \chi \bar{\chi} \phi(n_1,n_2) + 
c.c.)
\label{cubic}
\end{equation}
where $A_{\mu}(n_1,n_2)$ for $n_1=n_2=0$ is the $U(1)$ gauge field of the 
$p$-brane and for non-zero $n_1$ or $n_2$ is the massive vector bosons 
carrying the winding numbers $(n_1,n_2)$ with scalar partners 
$\phi(n_1,n_2)$.  Note that for zero winding
sector the coupling $C(0,0)$ to the $U(1)$ gauge field is 1 and it measures 
the charge of the tachyon with respect to this $U(1)$.

Now let us consider the other limit $x \rightarrow 1$. The intermediate
state now is the open string stretched between the $(p+2)$-brane carrying
the magnetic field to itself. Now $\tau^{-1} \rightarrow
-2\pi^2\sin^2(\epsilon)\log[(1-x)/\delta]$ goes to zero. Thus, in order
to get an interpretation of intermediate states, we must do a Poisson
resummation over the integers $n_1$ and $n_2$. The result is
\begin{equation}
I(x) = p_1.p_3 x^{s-1} (1-x)^{t-1} \frac{2\pi^3\sin(\epsilon)}{VF(1-x)}
\sum_{n_1,n_2} 
e^{-\frac{\pi^2}{\tau}(\frac{n_1^2}{ R_1^2} + \frac{n_2^2}{ R_2^2})}
\label{II}
\end{equation}
From the asymptotic behavior given above, we conclude that 
 $I(x)\rightarrow \frac{1}{N}p_1.p_3 (1-x)^{t-1} \\
 (\frac{(1-x)}{\delta})^{\frac{1}{2\pi^2N^2} (n_1^2 R_2^2 +
n_2^2 R_1^2)}$.
The $1/N^2$ in the exponent can be understood by considering the
$(p+2)$-$(p+2)$ open string with the modified boundary conditions due to the
magnetic field. Indeed one can easily work out the zero mode contribution to
the annulus partition function for this sector with the result that
effectively radii $R_1$ and $R_2$ are scaled by $1/N$. For zero windings
$n_1$ and $n_2$, if one integrates the above behaviour of $I(x)$ near $x=1$,
one finds a pole  $p_1.p_3/t$, with coefficient $1/N$. This pole is due to
the exchange of the $U(1)$ gauge field on the $(p+2)$-brane, and the
coefficient $1/N$ signifies that the corresponding charge of the tachyon is
$1/\sqrt{N}$. Just as in the $x\rightarrow 0$ case, we can identify the 
massive poles in the $t$-channel as arising from the following cubic terms
in the effective action: 
\begin{equation}
\sum_{n_1,n_2}(\tilde{C}(n_1,n_2) \bar{\chi}\partial_{\mu} \chi 
\tilde{A}_{\mu}(n_1,n_2)+c.c.) 
+\sum_{(n_1,n_2)\neq (0,0)}(\tilde{m} \tilde{C}(n_1,n_2) \chi \bar{\chi} 
\tilde{\phi}(n_1,n_2) + c.c.)
\label{cubic'}
\end{equation}
where $\tilde{C}(n_1,n_2) = \frac{1}{N}\delta^{-\frac{1}{4\pi^2N^2}(n_1^2 R_1^2 + 
n_2^2 R_2^2)}$, the mass $\tilde{m}=m/N$ and the fields $\tilde{A}$ and 
$\tilde{\phi}$ are the $(p+2)$-brane analogues of the fields $A$ and $\phi$.

To obtain the quartic term in the potential, we must remove the reducible
diagrams arising from the cubic terms in eqs.(\ref{cubic}) and 
(\ref{cubic'}). Including the exchanges of both the vector fields 
$A_{\mu}$'s and the scalar fields $\phi$'s in the $s$-channel and using 
the identity $(p_1-p_2).(p_3-p_4) = 4p_1.p_3 +s$ and similarly for the 
$t$-channel we find that the reducible diagrams contribute:
\begin{equation}
\sum_{n_1,n_2} \Bigl( (\frac{4p_1.p_3}{s+m^2} +1)|C(n_1,n_2)|^2 +
(\frac{4p_1.p_3}{t+\tilde{m}^2} +1)|\tilde{C}(n_1,n_2)|^2\Bigr)
\label{reducible}
\end{equation}
After removing
the reducible diagrams from the complete 4-pt. amplitude $\int_0^1 dx 
I(x)$, we can take $V\rightarrow 0$ limit, when the
tachyons become massless and the on-shell momenta can be set to zero.
The resulting quartic coupling is:
\begin{equation}
\sum_{n_1,n_2}e^{-\pi N(n_1^2 \frac{R_1}{R_2} + n_2^2\frac{R_2}{R_1})} +
\sum_{n_1,n_2}\frac{1}{N}e^{-\frac{\pi}{N}(n_1^2 \frac{R_1}{R_2} + 
n_2^2\frac{R_2}{R_1})} 
\label{quartic}
\end{equation}
This expression gives the quartic coupling $|\chi|^4$ in the effective 
action involving the infinite set of scalar fields $\phi(n_1,n_2)$ and
$\tilde{\phi}(n_1,n_2)$ whose mass squares are of order $V$.  
The minimization of this potential (including the cubic terms discussed 
above) results in an infinite set of equations for the scalar fields
$\phi(n_1,n_2)$, $\tilde{\phi}(n_1,n_2)$ and $\chi$. Solving the equations
for $\phi(n_1,n_2)$ and $\tilde{\phi}(n_1,n_2)$ one finds that to the 
leading order $\phi(n_1,n_2) = -\frac{C(n_1,n_2)}{m}|\chi|^2$ and 
similarly for $\tilde{\phi}$'s. Substituting this solution results in an 
effective quartic potential for $\chi$ with the quartic coupling being 
just the $n_1=n_2=0$ term of eq. (\ref{quartic}). 
This is the result we have used in section 2.

\end{document}